# Exchange-induced frustration in Fe/NiO multilayers

N. Rougemaille,[1,2] M. Portalupi,[3,4] A. Brambilla,[3] P. Biagioni,[3] A. Lanzara,[1,4] M. Finazzi,[3] A.K. Schmid,[1] and L. Duò[3]

[1]*Lawrence Berkeley National Laboratory, 1 Cyclotron Road, Berkeley, California 94720, USA*

[2]*Institut Néel, CNRS & Université Joseph Fourier, BP166, F-38042 Grenoble cedex 9, France*

[3]*L-NESS, Dipartimento di Fisica, Politecnico di Milano, piazza L. da Vinci 32, 20133 Milano, Italy*

[4]*Department of Physics, University of California Berkeley, Berkeley, California 94720, USA*

**Abstract**

Using spin-polarized low-energy electron microscopy to study magnetization in epitaxial layered systems, we found that the area vs perimeter relationship of magnetic domains in the top Fe layers of Fe/NiO/Fe(100) structures follows a power-law distribution, with very small magnetic domain cutoff radius (about 40 nm) and domain wall thickness. This unusual magnetic microstructure can be understood as resulting from the competition between antiferromagnetic and ferromagnetic exchange interactions at the Fe/NiO interfaces, rather than from mechanisms involving the anisotropy and dipolar forces that govern length scales in conventional magnetic domain structures. Statistical analysis of our measurements validates a micromagnetic model that accounts for this interfacial exchange coupling.

# I. INTRODUCTION

Magnetic domains and domain walls in ferromagnetic (FM) films grown on top of an antiferromagnetic (AFM) substrate often have different properties from those found in FM films grown on nonmagnetic substrates. For example, the FM domains observed in AFM/FM systems can be comparatively small; domain sizes smaller than 1 μm have been observed in some AFM/FM systems [1-8]. This is interesting, both because the performance potential of many spin-electronics concepts depends on the stability of small magnetic domains [9], and because one would like to understand the nature of the basic forces that govern the stability of magnetic domain structures in AFM/FM systems.

Minimal domain size is related to basic properties of the domain walls separating magnetic domains. Clearly, domains cannot be smaller than the width of the walls separating them. Approximately, one can often estimate the width δ of domain walls from the relation [9,10] $\delta \approx ab\sqrt{J_F / K}$ where $b$ is the atomic lattice spacing, $J_F$ is the strength of the exchange interaction (favoring greater wall width), $K$ is the magnetocrystal anisotropy energy (favoring smaller wall width), and the value of the factor $a$ depends on details of the spin structure of the domain wall, for example, $a \approx 10$ in bulk Fe [11]. Since $J_F >> K$ in most ferromagnetic materials, domain walls are often large compared to the atomic spacing. In bulk Fe, from the accepted values $J_F \approx 100$ meV and $K \approx 4$ μeV/atom [12-14], one expects that domain walls are a few hundred nanometer wide. By resolving the spin structure of FM thin films, spin-polarized low-energy electron microscopy [15] (SPLEEM) can be used to check whether this simple textbook picture is adequate to describe a material [16]. The SPLEEM image reproduced in Fig. 1(a) shows two large magnetic domains in an epitaxial Fe/MgO(100) film, separated by an ~200 nm wide domain wall: the observed magnetic microstructure is quantitatively consistent with the textbook picture, as well as with reports in the literature [17,18].

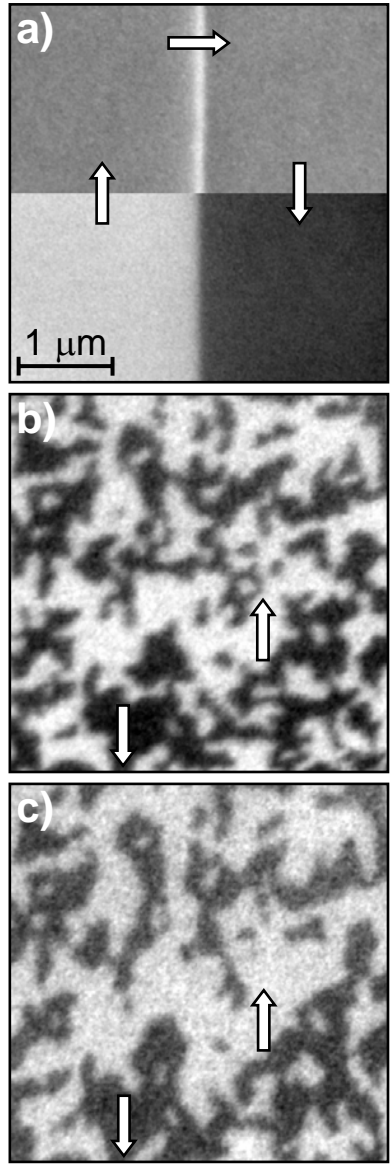

FIG. 1. 4×4 μm SPLEEM images of typical magnetic domains in (a) a Fe(400 nm)/MgO(100) film and Fe(*t*)/NiO(4.5 nm)/Fe(400 nm)/MgO(100) trilayers, with (b) *t*=1.3 nm and (c) *t*=6 nm, showing domain coarsening. Direction of magnetization (arrows) in bright vs dark domains is rotated by 180°. In panel (a), 180° domain walls are sufficiently wide ($\delta \approx 240$ nm) so that illumination with electron beam using 90° polarization reveals spin structure of the Néel wall [upper half of panel (a)].

In this work, we describe how SPLEEM imaging can be used to probe basic magnetic phenomena associated with AFM/FM coupling. SPLEEM images of the Fe capping layers in Fe/NiO/Fe(100) trilayer structures reveal magnetic domain structures with certain features, including the presence of extremely small domains (down to a few tens of nanometers), which suggest the presence of additional forces beyond the simple picture outlined above. We discuss how these observations can be understood on the basis of a recent model [19], and how analysis of the images can be used to estimate the strength of exchange forces at the AFM/FM interfaces. The premise of the model is that in AFM/FM systems, domains can be stabilized by frustration due to the interplay between the exchange interactions in the AFM and FM layers, and at the interfaces. By means of a statistical analysis of our images, we show that this model allows one to predict the correct dependence of the minimum domain size as a function of the Fe capping layer thickness.

## II. EXPERIMENTAL DETAILS

SPLEEM is an imaging technique that is based on the fact that the reflectivity of a magnetic surface for a spin-polarized low-energy electron beam depends on the relative orientation of the magnetization and the spin polarization of the illuminating electron beam [15]. The spin polarization of the electron beam can be oriented in all spatial directions, thus allowing the determination of unknown magnetic domain microstructures. Combining real-time imaging capability and good spatial resolution with very good magnetic sensitivity, SPLEEM is a uniquely suitable tool for quantitative characterization of magnetic configurations and structure-property relationships.

Our samples were grown on top of clean MgO(100) crystals, using well established [20] *in situ* preparation techniques. Low-energy electron diffraction (LEED) reveals that Fe/MgO(001) has an excellent crystal quality. NiO deposited on top of Fe/MgO(100) also grows epitaxially. Increased spot size and diffuse background suggest the presence of defects in the NiO layers, but the LEED patterns clearly indicate registry of the (100) direction of the NiO lattice parallel to the (110) direction of the Fe lattice. Freshly fabricated NiO/Fe/MgO(100) samples (thickness: MgO =bulk, Fe≈400 nm, and NiO≈4 nm) were transferred in ultrahigh vacuum into the specimen chamber of our SPLEEM, where deposition of the top Fe layer [21] (thickness, $t$=0–6 nm) and image acquisition took place at room temperature and under conditions of very low stray-magnetic field. Our samples were neither field cooled nor field grown, and therefore did not show macroscopic exchange bias [22].

## III. RESULTS AND DISCUSSION

### A. Topology of magnetic domains

In contrast to the large (tens or hundreds micrometer wide) magnetic domains separated by $\delta \sim 200$ nm thick domain walls with typically low curvature that we see in Fe/MgO(100) films [Fig. 1(a)], the magnetic domain patterns we observe in Fe capping layers of our Fe/NiO/Fe trilayer structures are strikingly different. The SPLEEM images reproduced in Figs. 1(b) ($t$=1.3 nm) and 1(c) ($t$ =6 nm) show intricate domain patterns with features that are unusual for films magnetized within the surface plane. The unusual features include many domains with surprisingly small sizes, topologically unconnected domains ("bubbles") and narrow elongated domains ("channels"). The domains are separated by narrow 180° domain walls with many high-curvature segments. The domain wall width in these Fe capping layers is close to the resolution limit of our instrument and we can only estimate an upper limit of $\delta$ <40 nm. Note how in many cases diameters of the entire domains are smaller than the widths of just the walls of domains in bare Fe/MgO(100) layers. The central question we address in this work is how one might understand the magnetic microstructure we find in our FM/AFM/FM trilayer samples.

It is interesting to note that the topologies of the magnetic domain patterns observed in Figs. 1(b) and 1(c) appear very similar to the predictions from two-dimensional Ising models with only short-range interactions [23-24]. For example, unconnected topologies with domain size distributions that include very small bubbles, as well as spanning domains that extend across the entire sample, are a natural feature of two-state systems [24]. What might seem puzzling is that if one neglects the magnetization of the substrate Fe layer, our system has fourfold symmetry and one might have guessed that the observed topologies should be closer to predictions of models with four degenerate ground states, such as four-state Potts models. Four-state models would predict topologies dominated by much more compact domain shapes, rare bubbles, and absence of spanning domains [25-26]. This is clearly not our observation here. Why do domains in the top

Fe layer "choose" one of the two degenerate easy axes and "ignore" the other one? This can be understood as a result of exchange coupling at the Fe/NiO interfaces. Essentially, the easy axis of magnetization in the ultrathin ferromagnetic Fe capping layer is coupled, through the antiferromagnetic NiO spacer [27-28], to the direction of the magnetization of the much thicker FM layer below. The thick FM substrate has very large magnetic domains [as shown in Fig. 1(a)], magnetized along either [001] or [010] directions. The antiferromagnetic easy axis of NiO grown on top of such a domain aligns with respect to the magnetization direction of the Fe layer below via exchange coupling. As a result, the spin structure within the NiO spacer scrambles the coupling direction with respect to the subsequently deposited top Fe layer, but conserves the easy axis with respect to the underlying domain in the Fe substrate. This exchange coupling, mediated by the NiO spacer layer, is the origin of uniaxial, in-plane magnetic anisotropy in the top Fe layer of our system. Here, we should point out that our prior studies of the magnetization reversal behavior in trilayer samples excludes the possibility that the coupling between the two Fe layers is due to ferromagnetic bridges associated with pinholes in the NiO spacers [29]. It is also known that, as a function of the NiO thickness, interesting situations can be achieved where magnetizations in the two Fe layers are aligned orthogonal to each other [29-30]. In imaging such structures (not shown here), we found that the topology of domains in the Fe capping is similar. Here, we limit our discussion to samples in which magnetizations in the upper and lower Fe layers are collinear. For this investigation, the only parameter we vary is the Fe capping layer thickness $t$. When $t$ is varied, the FM/AFM interfaces in our samples remain stable as far as their chemistry, morphology, and Fe-Fe coupling are concerned [29].

Within the conventional picture of domain walls in ferromagnetic material (see Introduction), an interpretation of the short defining length scales we observe is problematic. In the rest of this paper, we will explore the idea that the key toward understanding these domain patterns is interfacial exchange coupling. Exchange coupling is well known to play an important role at interfaces between NiO(100) and ferromagnetic layers [8,22,31-32], and we can test whether invoking exchange coupling leads to a plausible explanation of the observed magnetic domain structures. Models that have been proposed to explain the ubiquitous observation of exchange bias across AFM-FM interfaces include the suggested presence of uncompensated AFM spins associated with interface roughness [33], as well as noncollinear (or "spin-flop") coupling [34], and other models [22]. For our discussion, we do not need to choose one model. In any case, it is plausible to assume that on a microscopic length scale, the value of the interfacial exchange coupling fluctuates as a function of position, as a result of fluctuations of the detailed spin structure of the AFM layer. Random-field Ising models have been used to capture this type of disorder and, indeed, it appears that our multilayer FM/AFM system shares many of the general ground-state properties predicted for disordered Ising systems [24]. By analyzing the magnetic microstructure in the top Fe layer of our trilayers, we find that domains are fractal with fractal dimension 1.62, which is in good agreement with theoretical results from 2D Ising models [24]. Unless we are depositing additional Fe, we find no evidence of coarsening: the domain patterns are stable as a function of time, similar to the ground states of disordered Ising systems [24].

While Ising models are commonly applied to thin-film systems where magnetization is perpendicular to the surface plane, our Fe/NiO/Fe system is different in that the two states are in-plane magnetized. This distinction is important because in perpendicular magnetic films competition between the exchange interaction and dipolar forces can stabilize domain patterns on mesoscopic length scales [35]. In our case, dipolar interactions are not the driving forces that stabilize small domains and a different explanation is needed to understand the short defining length scales of these domain patterns.

One might have conjectured that accumulation of epitaxial strain during the fabrication process of our structures increases the magnetic anisotropy energy and thus leads to narrower domain walls. However, recalling the conventional view outlined above (i.e., $\delta \approx ab\sqrt{J_F / K}$ ), this leads to implausible conclusions: to explain the roughly tenfold reduction of domain wall width in our Fe capping layers (as compared to bulk Fe), the value of $K$ would have to have increased roughly by 2 orders of magnitude. A change of the anisotropy of this magnitude is not likely to occur due to magnetoelastic effects.

**B. Model**

In AFM-FM layered systems, domain wall widths are not necessarily determined, as in a bulk FM material, by the balance between magnetocrystal anisotropy and exchange energy. Rather, domain walls can be stabilized by frustration due to the interplay between the exchange interactions in the AFM and FM layers, and at the interfaces. One can appreciate an intuitive picture by considering uncompensated AFM-FM interfaces, where AFM atoms within atomic planes parallel to the surface have a net average magnetic moment. In that case, the average magnetic moment reverses direction at atomic-height steps and as a result, exchange coupling of the FM layer reverses direction on the two sides of the step. This frustration stabilizes domain walls near steps. Recent models describing this situation [19] suggest that domain walls can be much thinner in this type of structures than in bulk materials. NiO(100) is a nominally compensated AFM interface where, intuitively, one might have expected that the AFM spins pinning the FM layer cancel and interfacial exchange coupling would therefore vanish. However, this intuitive view is well known to be inconsistent with experimental observations, which show that the coupling strength is clearly greater than zero. In fact, most nominally compensated AFM surfaces usually exhibit substantial exchange bias, sometimes stronger than uncompensated surfaces of the same materials [22].

A relevant dimensionless parameter to capture the relative strength of exchange coupling is given by $\alpha = J_{int} S_{AF} / J_F S_F$, where $J_{int}$ is the exchange constant characterizing the interaction of spins belonging to different layers across the AFM/FM interfaces and $S_{AF}$ ($S_F$) the spin of the AFM (FM) atoms. We extend the ideas presented in Ref. 19 to include exchange coupling at nominally compensated surfaces, such as our case of NiO(100), by introducing a modified parameter $\alpha_z = z\alpha$. Here, we have used the same parameter $z$, $0 < z \leq 1$, that has been defined by Malozemof [36] to account for the reduced energy per unit area associated with the exchange coupling in a real sample, as compared to the theoretically limited case of an ideal, defect-free uncompensated surface. In fact, it is well documented that the strength of the average interfacial exchange coupling on a macroscopic area is usually found to be substantially smaller than exchange forces in bulk materials, typically of the order of $\alpha_z \approx 10^{-5} - 10^{-3}$ [22]. In an idealized microscopic model of uncompensated AFM interfaces, the value of $z$ might be unity [19]. In nominally compensated interfaces, $z$ might be the relative density of uncompensated AFM spins associated with interface roughness, if roughness is the source of exchange coupling [36].

We can then consider a critical radius ($r_{min}$) below which topologically unconnected FM bubbles collapse under the pressure due to the domain wall energy [24]. The value of $r_{min}$ depends on the balance between the energetic cost associated with the domain wall surrounding the bubble ($E_{DW}$), and the maximum energetic gain associated with the exchange coupling at the FM/AFM interface ($E_{F,AF}$). If a small, circular domain with radius $r$ covers the interfacial area $\pi r^2$, then the interfacial exchange coupling energy can be expressed as $E_{F,AF} = \pi (r/b)^2 z J_{int} S_{AF} S_F = \pi r^2 \alpha_z J_F S_F^2 / b^2$. This result is qualitatively different from what can be expected from Malozemoff's model [36]. The latter does not consider the domain hierarchy, the domain structure being a chessboard where all the domains have a lateral dimension equal to $r$. This leads to an average interface energy density scaling as $1/r$, i.e., to an interfacial exchange coupling energy on the domain footprint scaling as $r$ and not as $r^2$, as predicated by Ref. 24 for the minimum size of domains. An indication that the interfacial energy is indeed the mechanism that governs the minimum domain stability can be found in Ref. 29. In this work, we demonstrate that the reversal properties of the Fe capping in Fe/NiO/Fe trilayers are essentially determined by interface coupling rather than by volume defects, which influence domain nucleation barriers and might be important in determining the micromagnetic structure of thin FM films in other contexts [37].

To estimate $E_{DW}$, we can follow Ref. 19 and consider that the exchange energy per unit length of a domain wall in the FM capping layer is $w_F = J_F S_F^2 t / (b \sqrt{bt/\alpha_z})$. Furthermore, a domain wall in the Fe overlayer might extend into the NiO spacer thickness and maybe even in the Fe substrate [19], corresponding to an additional cost in frustrated exchange energy that we indicate as $w_0$, which does not depend on $t$. We therefore obtain $E_{DW} = 2\pi r (w_0 + w_F)$. According to our model, the minimal size of stable domains can be estimated by balancing $E_{DW}=E_{F,AF}$, which gives

$$r_{min} = 2w_0 b^2 / \alpha_z J_F S_F^2 + 2\sqrt{bt/\alpha_z} \qquad (1)$$

at variance with Malozemoff's model, which predicts a domain size scaling proportionally to the FM film thickness [36]. While a detailed estimate of the value of $w_0$ exceeds the scope of this paper, this result suggests that we can measure from our SPLEEM images the value of $r_{min}$ as a function of $t$. Confirming the square root dependence of $r_{min}$ on $t$, this allows us to determine an experimental estimate of the value of $\alpha_z$.

**C. Domain coarsening**

A qualitative analysis clearly shows that when we image the Fe/NiO/Fe trilayers during *in situ* deposition of the Fe capping layer, the Ising-like domain patterns coarsen as a function of capping layer thickness $t$ [see Figs. 1(b) and 1(c)]. Consistent with the ground-state properties of disordered Ising systems [24], we have not seen evidence of thermally activated coarsening of the domain structures. Note that the thickness dependence of the coarsening excludes that the FM domains in the top Fe layer are simply replication of the underlying AFM domain structure, as observed in other cases (see Ref. 7, for example). A common statistical probe to test domain growth is the pair correlation function $C(x,t) = \langle I(x+r,t)I(x,t) \rangle$, which measures the average of the product of intensities $I$ from pairs of image pixels separated by the distance $r$ as a function of the parameter $t$. In Fig. 2(a), we show a logarithmic plot of $C(x, t)$ from several SPLEEM images (one for each value of $t$), which indicates that the $C(x, t)$ functions are basically straight lines. The characteristic length scale of the domain patterns is therefore inversely proportional to the slope $S(t)$ of $C(x, t)$, taken at $x=0$. A plot of $1/S(t)$ vs $t$, as shown in Fig. 2(b), is consistent with the square-root dependence on $t$ that we expect from Eq. (1). One might argue that a square-root variation is not the most appropriate function to fit our measurements. For example, we could try to fit the data in Fig. 2(b) with a linear trend. However, in that case, we find that the standard deviation gets more than twice the value of the one found when a square-root trend is used. This is why we can reasonably exclude a linear trend.

To extract the value of $r_{min}$ from our images, we start by measuring the perimeter ($P$) versus area ($A$) relationships of the domains [24]. An example of our results is given in Fig. 2(c) for the case $t=1.3$ nm. The quantity $r_{min}$ can be estimated by plotting an additional line $A=P^2/4\pi$ to represent the $A$ versus $P$ relation of perfect circles. Since the $A/P$ ratio is maximum for circles, the intersection of the circle line with the line fitted to the domain measurements is our experimental estimate of $r_{min} \approx 40$ nm [24], which is indeed an unusually small cutoff radius for magnetic domains in continuous crystalline films.

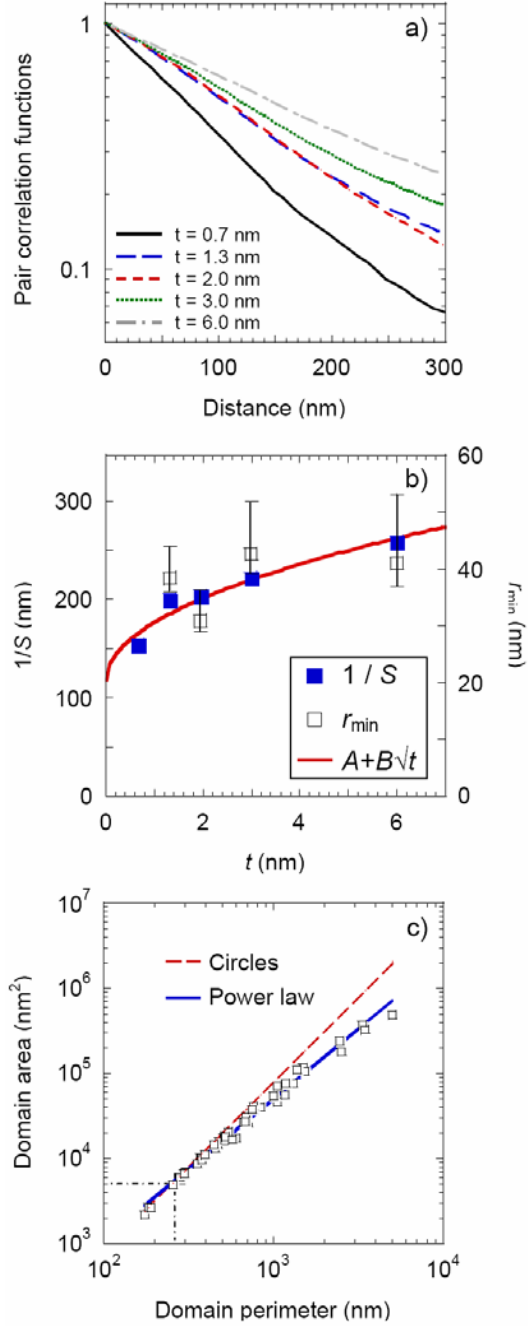

FIG. 2. (Color online) (a) Pair correlation profiles measured in our SPLEEM images for various values of the Fe capping layer thickness $t$. (b) $1/S$ [obtained from (a)] and $r_{min}$ [obtained for each intersection point obtained as shown in (c)] vs $t$. Error bars are indicated for $r_{min}$. Solid line: $(A+B\sqrt{t})$ best fit of both data sets which allows relative adjustment of the two vertical scales. (c) Area ($A$) vs perimeter ($P$) relationship for $t=1.3$ nm obtained on a large portion of the sample which includes the region shown in Figs. 1(b). Dashed line is the $A$ vs $P$ relationship for circles. Intersection corresponds to $r_{min} \approx 40$ nm.

Our measurement of the dependence of $r_{min}$ on the capping layer thickness $t$ can be interpreted in terms of interface coupling and domain wall energy density [24]. According to the scaling hypothesis [23], the length scales $r_{min}$ and $1/S$ are expected to follow the same variation with $t$. Clearly, the determination of the smallest domain size using the perimeter versus area relationship of the domains is affected by large error bars. This is so, mainly because only a limited number of domains is captured within the microscope field of view: for statistical reasons, $1/S$ can be measured with higher accuracy than $r_{min}$. Since we measure both quantities, it is convenient to first determine the appropriate proportionality factor to fit both measurements into one plot, as shown in Fig. 2(b) and then fitting Eq. (1) to the data. This allows us to estimate the value $\alpha_z = (1.1 \pm 0.3) \times 10^{-2}$. Taking into account the size of the NiO unit cell and estimating that the values of $J_F$ and $J_{int}$ are both of the order of 100 meV/atom, we find that the strength of the exchange coupling at these Fe/NiO interfaces is approximately 2.5 meV/nm$^2$. This value is roughly 1 order of magnitude larger than what one typically finds in macroscopic hysteresis loop measurements of NiO based exchange couples (for example, as summarized in Ref. 22). This difference further supports the importance of fluctuations in the spin structure of the AFM partner in exchange-biased systems [6]. The $r_{min}$ value we measure is governed by the stochastic distribution in the spin population of magnetic defects. Within local interface regions, the average spin direction of ensembles of magnetic defects does not average to zero, thus stabilizing small domains in the Fe capping layer [6,24]. On a larger scale that spans macroscopic samples, the tendency of magnetic defects to cancel, on average, is the reason why exchange bias is much weaker in macroscopic samples, even after field cooling.

## IV. CONCLUSION

In conclusion, we imaged the magnetic microstructure of thin Fe films epitaxially grown on NiO/Fe/MgO(100) and found unusually small domains, with magnetic domain cutoff radius down to 40 nm and domain wall thickness below the microscope resolution. Using a model that takes into account frustrated exchange interactions at FM/AFM interfaces, we reproduce the correct thickness dependence of domain coarsening and understand the microscopic mechanism which governs the stabilization of small domains in terms of fluctuations of the AFM spin structure at the interfaces. In light of potential applications, it is interesting to note that the stability of our observed nanodomains coincides with perfect remanence [29]. Magneto-optic Kerr measurements show that remanence of the films can be as large as saturation, which indicates that single-domain states spanning the sample are possible, and one might envision to "write" individual, stable "bubble" nanodomains into such multilayer structures. It is also interesting to note that high-performance spin-tunnel junctions have been realized with structures based on epitaxial Fe/MgO(100) interfaces [38-39]. Conceivably, the implementation of a NiO layer in future spin-tunneling devices based on Fe/MgO(100) epitaxial structures might enable a high degree of miniaturization.

## ACKNOWLEDGEMENTS

We wish to thank N. Bartelt and Q. Ramasse for helpful discussions, T. Ravazzani and M. Marcon for their help with the code for image recognition. This work was supported by the U.S. Department of Energy under Contract No. DEAC02-05CH11231 and under Contract No. DEAC03-76SF00098, by the Délégation Générale pour l'Armement under Contract No. 9860830051, and by the National Science Foundation through Grant No. DMR03-49361.